\documentstyle[epsf]{article}
\begin{document}
%%%%%%%%%%%%%%%%%%%%%%%%%%%%%%%%%%%%%%%%%%%%%%%%%%%%%%%%%%%%%%%%%%%%%
\title{
Large-$q$ Expansion of the Specific Heat 
for the Two-dimensional $q$-state Potts Model}
\author{
        H. ARISUE \\
        Osaka Prefectural College of Technology, \\
        Saiwai-cho, Neyagawa, Osaka 572, Japan
              \and
        K. TABATA \\
        Osaka Institute of Technology, Junior College, \\
        Ohmiya, Asahi-ku, Osaka 535, Japan}
\maketitle
\begin{abstract}
We have calculated the large-$q$ expansion for the specific heat at the phase 
transition point in the two-dimensional $q$-state Potts model to the 23rd 
order in $1/\sqrt{q}$ using the finite lattice method. The obtained series 
allows us to give highly convergent estimates of the specific heat for $q>4$ 
on the  first order transition point. The result confirm us the correctness of 
the conjecture by Bhattacharya {\em et al.} on the asymptotic behavior of the 
specific heat for $q \rightarrow 4_+$. 
\end{abstract}
%%%%%%%%%%%%%%%%%%%%%%%%%%%%%%%%%%%%%%%%%%%
\newpage
The $q$-state Potts model on the square lattice 
has a first order phase transition for $q>4$.
The phase transition point and many quantities 
at the transition point are known exactly, including the 
free energy, the internal energy\cite{Baxter1973} and the correlation 
length\cite{Klumper}.
On the other hand, important quantities such as the specific heat 
and the susceptibility are not solved exactly.
The correlation length increases to infinity as $q\rightarrow 4_+$,
so it becomes very difficult in general to evaluate 
the quantities that are not solved exactly
for $q$ as `small' as 5,
where the correlation length reaches as large as several thousands.

Here we  concentrate on the specific heat at the transition point. 
Many methods have been used to evaluate it including the Monte Carlo 
simulations\cite{Janke}, the low-(and high-)temperature 
expansion\cite{Lowtemp}, and the large-$q$ expansion\cite{Bhattacharya1997}, 
among which the large-$q$ expansion appears very promising.
Bhattacharya {\em et al.} calculated the large-$q$ series for the specific heat
and higher energy cumulants to order 10 in $z\equiv 1/\sqrt{q}$.
Making the Pad\'e analysis of the series assuming the asymptotic form 
in the limit of $q\rightarrow 4_{+}$, 
they obtained very precise values of the specific heat for $q\ge 7$,
whose accuracy seems to be higher than the Monte Carlo simulations.

In this paper we will enlarge the large-$q$ series for the specific heat
at the transition point to order 23 in $z$.
We use the finite lattice method\cite{Enting1977,Creutz,Arisue1984}, 
which has been used mainly to 
generate the low- and high-temperature series in statistical systems and
the strong coupling series in lattice gauge theory.
The method can give in general longer series 
than those generated by the graphical method in lower space (and time) 
dimensions. 
In the graphical method, 
one has to list up all the relevant diagrams and count the number they appear.
In the finite lattice method we can skip the job and 
reduce the main work to the calculation of the expansion of the partition 
function for a series of finite size lattices,
which can be done using the straightforward site-by-site 
integration\cite{Entingtwo} without the graphical technique.

The model is defined by the partition function
\begin{equation}
  Z=\sum_{\{s_i\}} \exp{(-\beta H)},  \qquad  
  H=-\sum_{\langle i,j \rangle}\delta_{s_i,s_j}\;,
\end{equation}
where $\langle i,j \rangle$ represents the pair of nearest neighbor sites 
and $s_i=1,2,\cdots,q$.
The phase transition point $\beta_t$ is given by $\exp{(\beta_t)-1=\sqrt{q}}$.
We will consider the free energy density in the disordered phase, 
which is given by 
\begin{equation}
F(\beta)_d = \lim_{L_x,L_y\rightarrow \infty}(L_x L_y)^{-1}\ln{Z(L_x,L_y)}\;, 
\label{eq:Free_energy}
\end{equation}
where the partition function for the $L_x\times L_y$ lattice should be 
calculated 
with the free boundary condition corresponding to the disordered phase.
The large-$q$ expansion of the partition function can be given through the 
Fortuin-Kasteleyn representation\cite{Fortuin} as
\begin{eqnarray}
    Z(L_x,L_y)
      &=& q^{L_x L_y}\sum_{l,j} N_{l,j}\left(e^\beta -1\right)^{l} q^{-j}\; 
      \nonumber\\
      &=& q^{L_x L_y}\sum_{l,j} N_{l,j}Y^{l} z^{2j-l}\;, 
\end{eqnarray}
where  $N_{l,j}$ is the number of configurations of $l$ bonds 
connecting the nearest neighbor sites on the $L_x \times L_y$ lattice
with $L_x L_y-j$ independent clusters of sites 
and $Y\equiv (e^{\beta}-1)/\sqrt{q}$. 
(Two sites connected to each other belong to the same cluster.)

We define $H(l_x,l_y)$
for each $l_x \times l_y$ lattice ($l_x,l_y=1,2,3,\cdots$) as\cite{Arisue1984}
\begin{equation}
   H(l_x,l_y) = \ln{ [Z(l_x,l_y)/q^{l_x l_y}]}\;,
\end{equation}
where $Z(l_x,l_y)$ is the partition function with the free boundary condition, 
and define $W(l_x,l_y)$ recursively as
\begin{eqnarray}
&&   W(l_x,l_y) = H(l_x,l_y) \nonumber\\
&& \qquad - \sum_{l_x^{\prime}\le l_x,l_y^{\prime}\le l_y}^{\ \ \ \ ({\prime})}
        (l_x-l_x^{\prime}+1)(l_y-l_y^{\prime}+1)W(l_x^{\prime},l_y^{\prime})\;.
\end{eqnarray}
Here the $(')$ indicates that a term
with $l_x^{\prime}=l_x$ and $l_y^{\prime}=l_y$ should be excluded in the 
summation.
Then the free energy density defined by Eq.(\ref{eq:Free_energy}) is given by
\begin{equation}
    F(\beta)_d = \ln{(q)}+\sum_{l_x,l_y} W(l_x,l_y)\;.
\end{equation}

 We can prove 
\cite{Arisue1984} 
that the Taylor expansion of the $W(l_x,l_y)$ 
with respect to $z$ and $Y$ includes 
the contribution from all the clusters of polymers 
in the standard cluster expansion 
that can be embedded into the $l_x \times l_y$ lattice 
but cannot be embedded into any of its rectangular sub-lattices.
Each cluster that contributes to the lowest order term
of the $W(l_x,l_y)$ consists of a single polymer
and it has the order of $z^{l_x+l_y-2}$. 
An example of such a single polymer is 
shown in Fig. 1.
Therefore to obtain the series to order $z^N$
we have only to take into account all the rectangular lattices
that satisfy $ l_x+l_y-2 \le N$.
If we set $Y=1+y$\cite{Guttmann1993}, 
then we have only to keep the expansion with respect to $y$ to order $y^n$ 
to obtain the $n$-th energy cumulant at the phase transition 
point as,
\begin{equation}
    F_d^{(n)}=\left.\frac{d^n}{d\beta^n} F(\beta)_d \right|_{\beta=\beta_t}
             =\sum_m a^{(n)}_m z^m\;.
\end{equation}
(We note that $\frac{d}{d\beta}=(1+y+z)\frac{d}{dy}$
and $y=0$ at $\beta=\beta_t$.)
We can also calculate the series for the energy cumulant $F_o^{(n)}$ 
at $\beta_t$
in the ordered phase by using the duality relation
\begin{equation}
    F_d^{(2)}-F_o^{(2)}= -z [F_d^{(1)}-F_o^{(1)}]\;. \label{eq:Duality}
\end{equation}
We have calculated the series to order $N=23$ in $z$ for $n=0,1$ and $2$.
The obtained series for the zeroth and first cumulants
(i.e. the free energy and the internal energy) agree
with the expansion of the exactly known expressions.
The series for the second cumulants are listed in Table 1.
The coefficients for $F_o^{(2)}$ agree with those
by Bhattacharya {\em et al.} to order 10.

The latent heat ${\cal L}$ is known\cite{Baxter1973} to vanish 
at $q\rightarrow 4_+$ as
\begin{equation}
{\cal L} \sim 3\pi x^{-1/2}\;, \label{eq:asymptotic}
\end{equation}
with $x=\exp{(\pi^2/2\theta)}$ and $2\cosh{\theta}=\sqrt{q}$.
Bhattacharya {\em et al.}\cite{Bhattacharya1994}
made the conjecture that  $F_{d,o}^{(2)}$ will diverge
at $q\rightarrow 4_+$ as
\begin{equation}
F_{d,o}^{(2)} \sim \alpha x\;. \label{eq:conjecture}
\end{equation}
The constant $\alpha$ should be common for the ordered and disordered phases
from Eqs. (\ref{eq:Duality}) and (\ref{eq:asymptotic}). 
Here we follow this conjecture, then
the product $F^{(2)} {\cal L}^2 $ can be expected to be a smooth function 
of $\theta$, so we
apply the Pad\'e approximation 
to this quantity as
\begin{eqnarray}
F_d^{(2)} {\cal L}^2 &=& z P_M(z)/Q_L(z) + O(z^{M+L+2})\;, \nonumber\\ 
F_o^{(2)} {\cal L}^2 &=& z^2 R_M(z)/S_L(z) + O(z^{M+L+3})\;,
\end{eqnarray}
where $P_M(z)$ and $Q_L(z)$ ($R_M(z)$ and $S_L(z)$) are the $M$-th and
$L$-th order polynomials with $M+L+1\le N$ ($M+L+2\le N$).
We give in Table 2 the values of the 
specific heat $C_{d,o}$ evaluated from these Pad\'e approximants 
for some values of $q$ 
and present in Fig. 2 the behavior of the ratio of the $F^{(2)}$ to $x$ 
plotted versus $\theta$.
The averages and errors are taken from all the $[M,L]$ Pad\'e approximants
with $M\ge 8$ and $L\ge 8$, excluding that whose denominator has zero 
at some point in $4<q<\infty$. 
We have checked that the duality relation (\ref{eq:Duality}), which is not 
respected exactly by the Pad\'e approximants, is really 
satisfied within the accuracy for all the range of $q>4$. 
These estimates are more precise by three or four orders of magnitude than
(and of course  consistent with)
the previous result for $q\ge 7$ 
from the large-$q$ expansion to order $z^{10}$
by Bhattacharya {\em et al.}\cite{Bhattacharya1997} and 
the result for $q=10,15,20$
from the Monte Carlo simulations carefully done 
by Janke and Kappler\cite{Janke}. 
What should be emphasized is that we obtained the values of
the specific heat in the accuracy of about 0.1 percent
at $q=5$ where the correlation length is as large as 2500\cite{Klumper}.
As for the asymptotic behavior of $F^{(2)}$ at $q\rightarrow 4_+$,
the Pad\'e data of $F_d^{(2)}/x$ and $F_o^{(2)}/x$ have 
relatively large errors of a few percent arround $q=4$, 
but their behaviors shown in Fig. 2 are enough to convince us 
that the conjecture (\ref{eq:conjecture}) is true with
\begin{equation}
     \alpha = 0.073 \pm 0.002\;.
\end{equation}

The extension of the large-$q$ expansion to the higher energy cumulants
and the magnetization cumulants is rather straightforward and now in progress.

\clearpage
%%%%%%%%%%%%%%%%%%%%%%%%%%%%%%%%%%%%%%%%%%%%%%%%%%%%%%%%%%%
\begin{table}[htb]
\caption{
The large-q expansion coefficients $a^{(2)}_m$
for the second energy cumulant.
         }
\label{tab:coeff}
\begin{center}
\begin{tabular}{rrr}
\hline
    $m$  & \multicolumn{1}{c}{$a^{(2)}_m$(disordered)}
         & \multicolumn{1}{c}{$a^{(2)}_m$(ordered)}  \\
\hline
$  0$ &         0 & $         0$  \\ 
$  1$ &         2 & $         0$  \\ 
$  2$ &        14 & $        16$  \\ 
$  3$ &        26 & $        34$  \\ 
$  4$ &       118 & $       114$  \\ 
$  5$ &       250 & $       254$  \\ 
$  6$ &       894 & $       882$  \\ 
$  7$ &      1936 & $      1944$  \\ 
$  8$ &      6160 & $      6128$  \\ 
$  9$ &     13538 & $     13550$  \\ 
$ 10$ &     39774 & $     39698$  \\ 
$ 11$ &     88360 & $     88360$  \\ 
$ 12$ &    245188 & $    245036$  \\ 
$ 13$ &    547468 & $    547356$  \\ 
$ 14$ &   1457976 & $   1457784$  \\ 
$ 15$ &   3264012 & $   3263316$  \\ 
$ 16$ &   8410284 & $   8410596$  \\ 
$ 17$ &  18868858 & $  18865590$  \\ 
$ 18$ &  47391870 & $  47395762$  \\ 
$ 19$ & 106180532 & $ 106166828$  \\ 
$ 20$ & 261607968 & $ 261629456$  \\ 
$ 21$ & 586199668 & $ 586145660$  \\ 
$ 22$ &1415497756 & $1415594740$  \\ 
$ 23$ &3174285456 & $3174081000$  \\ 
\hline
\end{tabular}
\end{center}
\end{table}

%%%%%%%%%%%%%%%%%%%%%%%%%%%%%%%%%%%%%%%%%%%%%%%%%%%%%%%%%%%
\begin{table}[htb]
\caption{
The specific heat for some values of $q$.
         }
\label{tab:f(2)}
\begin{center}
\setlength{\tabcolsep}{1.5pc}
\newlength{\digitwidth} 
\settowidth{\digitwidth}{\rm 0}
\catcode`?=\active \def?{\kern\digitwidth}
\begin{tabular}{rll}
\hline
    $q$  & \multicolumn{1}{c}{$C_d$}
         & \multicolumn{1}{c}{$C_o$}  \\
\hline
$  5$ & 2889(2)            & 2886(3)             \\ 
$  6$ & ?205.93(3)         & ?205.78(3)          \\ 
$  7$ & ??68.738(2)        & ??68.513(2)         \\ 
$  8$ & ??36.9335(3)       & ??36.6235(3)        \\ 
$  9$ & ??24.58761(8)      & ??24.20344(7)       \\ 
$ 10$ & ??18.38543(2)      & ??17.93780(2)       \\ 
$ 12$ & ??12.401336(3)     & ??11.852175(2)      \\ 
$ 15$ & ???8.6540358(4)    & ???7.9964587(2)     \\ 
$ 20$ & ???6.13215967(2)   & ???5.36076877(1)    \\ 
$ 30$ & ???4.2989934145(6) & ???3.4128952554(3)  \\ 
\hline
\end{tabular}
\end{center}
\end{table}
\clearpage
%%%%%%%%%%%%%%%%%%%%%%%%%%%%%%%%%%%%%%%%%%%%%%%%%%%%%%%%%%%
\begin{figure}[h]
\caption{An example of the cluster consisting of a single polymer 
that contributes to the lowest order term of the $W(l_x,l_y)$ 
with $l_x=4$ and $l_y=5$.
The closed circles are the sites, and the solid lines and the crosses are
the bonds connecting and disconnecting the nearest neighbor sites, 
respectively, in the Fortuin-Kasteleyn representation.
}
\label{fig1}
\hspace{30mm}
\epsffile{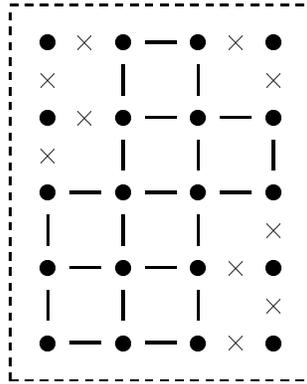}
\end{figure}
%%%%%%%%%%%%
\begin{figure}[h]
\caption{The ratio of the $F^{(2)}$ to $x$ 
plotted versus $\theta$.
The dashed and dotted lines represent the errors for the 
ordered and disordered ones, respectively.
}
\label{fig2}
\vspace{10mm}
\epsffile{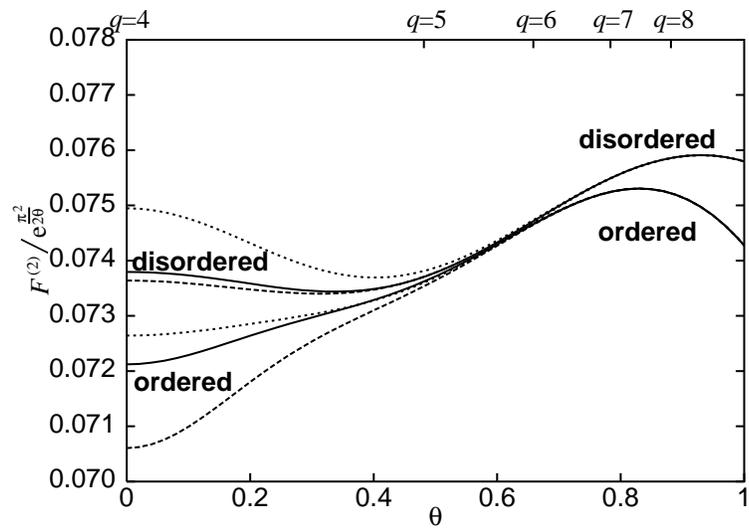}
\end{figure}

%%%%%%%%%%%%%%%%%%%%%%%%%%%%%%%%%%%%%%%%%%%%%%%%%%%%
\end{document}